\documentclass[12pt,letterpaper]{article} 

\usepackage{amsmath,amsfonts,amsthm,amssymb,stmaryrd,graphicx,relsize,bm}
\theoremstyle{plain}

\numberwithin{equation}{section}
\newtheorem{thm}{Theorem}[section]
\newtheorem{lem}[thm]{Lemma}

\allowdisplaybreaks  

{\begin{flushleft}\textbf{Example #1}.\enspace}%
{\end{flushleft}}

\newcounter{cond}

\newcommand{\integers}{{\mathbb Z}}

\newcommand{\gscript}{{\mathcal G}}
\newcommand{\bscript}{{\mathcal B}}
\newcommand{\cscript}{{\mathcal C}}
\newcommand{\lscript}{{\mathcal L}}
\newcommand{\qscript}{{\mathcal Q}}
\newcommand{\sscript}{{\mathcal S}}
\newcommand{\bbar}{\overline{b}}
\newcommand{\cbar}{\overline{c}}
\newcommand{\dbar}{\overline{d}}
\newcommand{\qbar}{\overline{q}}
\newcommand{\sbar}{\overline{s}}
\newcommand{\tbar}{\overline{t}}
\newcommand{\ubar}{\overline{u}}
\newcommand{\bhat}{\widehat{b}}
\newcommand{\chat}{\widehat{c}}
\newcommand{\shat}{\widehat{s}}

\newcommand{\ctimes}{\mathrel{\mathlarger\cdot}}

\newcommand{\doubleab}[1]{\left|\left|#1\right|\right|}
\newcommand{\brac}[1]{\left\{#1\right\}}
\newcommand{\paren}[1]{\left(#1\right)}
\newcommand{\sqbrac}[1]{\left[#1\right]}

\begin{document}

\title{THE ELEMENTARY PARTICLE CUBE
}
\author{S. Gudder\\ Department of Mathematics\\
University of Denver\\ Denver, Colorado 80208, U.S.A.\\
sgudder@du.edu
}
\date{}
\maketitle

\begin{abstract}
Postulating that spacetime is discrete, we assume that physical space is described by a 3-dimensional cubic lattice. The corresponding symmetry group of rotations has order 24 and motivates the introduction of a cubic shaped graph with 27 vertices and 351 edges. We call this graph the elementary particle cube (EPC) and consider the vertices as tiny cells that pre-elementary particles called preons can occupy and the edges as interactions between preons. The 23 nontrivial members of the symmetry group naturally associate with the 23 basic elementary particles. We assume that each elementary particle is described by a unique subgraph of the EPC. The particular subgraph is determined by symmetry and the particle's mass. We postulate that the particle mass is a certain function of the lengths of the edges in the graph representing the particle. This correspondence between particle graphs and mass appears to be quite accurate and gives a reason why leptons and quarks come in three generations. In this way, the basic elementary particles emerge in a natural way from a few simple principles. The paper ends with a discussion of hadrons, which as in the standard model, are composite systems of quarks.
\end{abstract}

\section{Introduction}  
This article presents an approach for studying elementary particles. Theoretical and experimental investigations in physics have identified and measured properties of the basic elementary particles. These particles have been organized into certain categories; namely, the force mediating bosons consisting of the photons, gravitons, gluons and weak bosons $W^-$, $Z^0$, the leptons consisting of the electron $e$, muon $\mu$, tau $\tau$ and their corresponding neutrinos $\nu_e, \nu _\mu, \nu _\tau$, the quarks, down $d$, up $u$, strange $s$, charm $c$, bottom $b$, top $t$ and finally, the Higgs boson $H$. In this article we shall mainly consider the 23 particles consisting of the gluons, weak bosons, leptons, quarks and Higgs. Their corresponding anti-particles need not be considered because they have identical properties except for charge when electric charge is present.

Our basic assumption is that spacetime is discrete \cite{bdp16,cro16,gud68,hei30,rus54}. We shall not be concerned with time here, so we assume that space has the structure of a 3-dimensional cubic lattice $\sscript _3=\integers ^3$ where $\integers$ is the set of integers. Other discrete configurations are possible \cite{gud161,gud162}, but we choose $\sscript _3$ because of its simplicity. The distance between adjacent lattice points of $\sscript _3$ are presumably on a Planck scale of about $10^{-33}$cm \cite{hei30,rus54}. We endow $\sscript _3$ with the usual Euclidean norm given by $\doubleab{v}=\sqrt{a^2+b^2+c^2}$ for $v=(a,b,c)\in\sscript _3$. This provides the metric $d(u,v)=\doubleab{u-v}$ on $\sscript _3$. In this way $(\sscript _3,\doubleab{\ctimes})$ becomes a normed module over $\integers$. The \textit{symmetry group} $\gscript '_3$ on $\sscript _3$ is the group of linear operators
($3\times 3$ matrices) $T\colon\sscript _3\to\sscript _3$ that satisfy $\doubleab{Tv}=\doubleab{v}$ for all $v\in\sscript _3$. The group of \textit{physical rotations} $\gscript _3$ is the subgroup of $\gscript '_3$ consisting of operators with unit determinant. The group $\gscript '_3$ has order 48 and $\gscript _3$ has order 24 \cite{gudap1,gudap2}. We shall show in Section~2 that the 23 nontrivial elements of $\gscript _3$ congregate into four natural subsets that we call the lepton, quark, gluon and boson types. As they should, these types contain 6,6,8,3 elements, respectively. The rotation axes of the various types suggest the construction of a cubic shaped graph with 27 vertices and 351 edges. We call this graph the elementary particle cube (EPC) and consider the vertices as tiny cells that pre-elementary particles called preons can occupy and the occupied edges as interactions between preons.

In Sections~3 and 4 we assume that each elementary particle is described by a unique subgraph of the EPC. The particular subgraph is determined by symmetry and the particle's mass. We propose a mass formula that gives the mass of a particle as a function of the lengths of the edges in the graph representing the particle. This correspondence between particle graphs and mass appears to be quite accurate. In particular, predicted particle masses agree with experiment to within about $1\%$ accuracy. The paper ends with a discussion of hadrons, which as in the standard model, are composite systems of quarks.

\section{Symmetry}  
As discussed in Section~1, we have the group of symmetries $\gscript '_3$ and the subgroup of physical rotations $\gscript _3$ on
$\sscript _3$. It is easy to check that $\gscript '_3=\gscript _3\cup (-\gscript _3)$. That is, every element $Z\in\gscript '_3$ satisfies $Z\in\gscript _3$ or $-Z\in\gscript _3$. We denote the elements of $\gscript _3$ by $A,B,\ldots,X$. These are orthogonal $3\times 3$ matrices with entries $0,1,-1$ and are given by \cite{gudap2}
\begin{align*}
&\hskip -10pt A=\begin{bmatrix}\noalign{\smallskip}
1&0&0\\0&0&-1\\0&1&0\\\noalign{\smallskip}\end{bmatrix},\ 
B=\begin{bmatrix}\noalign{\smallskip}
0&0&1\\0&1&0\\-1&0&0\\\noalign{\smallskip}\end{bmatrix},\ 
C=\begin{bmatrix}\noalign{\smallskip}
0&-1&0\\1&0&0\\0&0&1\\\noalign{\smallskip}\end{bmatrix},\ 
D=\begin{bmatrix}\noalign{\smallskip}
1&0&0\\0&-1&0\\0&0&-1\\\noalign{\smallskip}\end{bmatrix}\\\noalign{\medskip}
&\hskip -10pt E=\begin{bmatrix}\noalign{\smallskip}
1&0&0\\0&0&1\\0&-1&0\\\noalign{\smallskip}\end{bmatrix},\ 
F=\begin{bmatrix}\noalign{\smallskip}
-1&0&0\\0&1&0\\0&0&-1\\\noalign{\smallskip}\end{bmatrix},\ 
G=\begin{bmatrix}\noalign{\smallskip}
0&0&-1\\0&1&0\\1&0&0\\\noalign{\smallskip}\end{bmatrix},\ 
H=\begin{bmatrix}\noalign{\smallskip}
-1&0&0\\0&-1&0\\0&0&1\\\noalign{\smallskip}\end{bmatrix}\\\noalign{\medskip}
&\hskip -10pt I=\begin{bmatrix}\noalign{\smallskip}
1&0&0\\0&1&0\\0&0&1\\\noalign{\smallskip}\end{bmatrix},\ 
J=\begin{bmatrix}\noalign{\smallskip}
0&1&0\\-1&0&0\\0&0&1\\\noalign{\smallskip}\end{bmatrix},\ 
K=\begin{bmatrix}\noalign{\smallskip}
0&0&1\\1&0&0\\0&1&0\\\noalign{\smallskip}\end{bmatrix},\ 
L=\begin{bmatrix}\noalign{\smallskip}
0&1&0\\0&0&-1\\-1&0&0\\\noalign{\smallskip}\end{bmatrix}\\\noalign{\medskip}
&\hskip -10pt M\!=\!\begin{bmatrix}\noalign{\smallskip}
0&-1&0\\0&0&-1\\1&0&0\\\noalign{\smallskip}\end{bmatrix}\!,\,
N\!=\!\begin{bmatrix}\noalign{\smallskip}
0&-1&0\\0&0&1\\-1&0&0\\\noalign{\smallskip}\end{bmatrix}\!,\,
O\!=\!\begin{bmatrix}\noalign{\smallskip}
-1&0&0\\0&0&1\\0&1&0\\\noalign{\smallskip}\end{bmatrix}\!,\,
P\!=\!\begin{bmatrix}\noalign{\smallskip}
-1&0&0\\0&0&-1\\0&-1&0\\\noalign{\smallskip}\end{bmatrix}\\\noalign{\medskip}
&\hskip -10pt Q\!=\!\begin{bmatrix}\noalign{\smallskip}
0&0&-1\\-1&0&0\\0&1&0\\\noalign{\smallskip}\end{bmatrix}\!,\,
R\!=\!\begin{bmatrix}\noalign{\smallskip}
0&0&1\\0&-1&0\\1&0&0\\\noalign{\smallskip}\end{bmatrix}\!,\,
S\!=\!\begin{bmatrix}\noalign{\smallskip}
0&1&0\\1&0&0\\0&0&-1\\\noalign{\smallskip}\end{bmatrix}\!,\,
T\!=\!\begin{bmatrix}\noalign{\smallskip}
0&0&-1\\0&-1&0\\-1&0&0\\\noalign{\smallskip}\end{bmatrix}\\\noalign{\medskip}
&\hskip -10pt U\!=\!\begin{bmatrix}\noalign{\smallskip}
0&-1&0\\-1&0&0\\0&0&-1\\\noalign{\smallskip}\end{bmatrix}\!,\,
V\!=\!\begin{bmatrix}\noalign{\smallskip}
0&0&1\\-1&0&0\\0&-1&0\\\noalign{\smallskip}\end{bmatrix}\!,\,
W\!=\!\begin{bmatrix}\noalign{\smallskip}
0&1&0\\0&0&1\\1&0&0\\\noalign{\smallskip}\end{bmatrix}\!,\,
X\!=\!\begin{bmatrix}\noalign{\smallskip}
0&0&-1\\1&0&0\\0&-1&0\\\noalign{\smallskip}\end{bmatrix}\\
\end{align*}
We now show that various elements of $\gscript _3$ band together to form subsets with common properties. These common properties are order and invariant vectors. We call the set
\begin{equation*}
\lscript =\brac{A,B,C,E,G,J}
\end{equation*}
the \textit{lepton type}, the set
\begin{equation*}
\qscript =\brac{O,P,R,S,T,U}
\end{equation*}
the \textit{quark type}, the set
\begin{equation*}
\gscript =\brac{K,L,M,N,Q,V,W,X}
\end{equation*}
the \textit{gluon type} and the set
\begin{equation*}
\bscript =\brac{D,F,H}
\end{equation*}
the \textit{boson type}. It is easy to check that the elements of $\lscript$ have order 4, the elements of $\qscript$ and $\bscript$ have order 2 and the elements of $\gscript$ have order 3.

\textit{Invariant vectors} are eigenvectors with corresponding eigenvalue~1. To study such vectors, it is convenient to introduce the following elements of $\sscript _3$.
\begin{list}
{\textit{Level\ }\arabic{cond}:}{\usecounter{cond}
\setlength{\rightmargin}{\leftmargin}}
\item $u_1=(0,0,1), u_2=(0,0,-1), u_3=(0,1,0), u_4=(0,-1,0)$,\newline\hglue 16pt $u_5=(1,0,0), u_6=(-1,0,0)$
\item $v_1=(0,1,1), v_2=(0,1,-1), v_3=(0,-1,1), v_4=(0,-1,-1)$,
\newline\hglue 16pt $v_5=(1,0,1), v_6=(1,0,-1), v_7=(-1,0,1), v_8=(-1,0,-1)$,
\newline\hglue 16pt $v_9=(1,1,0), v_{10}=(1,-1,0), v_{11}=(-1,1,0), v_{12}=(-1,-1,0)$
\item $w_1=(1,1,1), w_2=(1,-1,1), w_3=(1,1,-1), w_4=(-1,1,1)$,
\newline\hglue 16pt $w_5=(-1,-1,1), w_6=(-1,1,-1), w_7=(1,-1,-1)$,
\newline\hglue 16pt $w_8=(-1,-1,-1)$
\end{list}
The reason for the level terminology is because $\doubleab{u_j}^2=1$, $\doubleab{v_j}^2=2$, $\doubleab{w_j}^2=3$ for all applicable $j$. We do not normalize these vectors because we want them to be in $\sscript _3$.

Let us find the invariant vectors for some elements of $\gscript _3$. For $A$ we have
\begin{equation*}
Au_5=\begin{bmatrix}\noalign{\smallskip}
1&0&0\\0&0&-1\\0&1&0\\\noalign{\smallskip}\end{bmatrix}\ 
\begin{bmatrix}\noalign{\smallskip}1\\0\\0\\\noalign{\smallskip}\end{bmatrix}=
\begin{bmatrix}\noalign{\smallskip}1\\0\\0\\\noalign{\smallskip}\end{bmatrix}=u_5
\end{equation*}
Also, $Au_6=u_6$ so $A$ has two level~1 invariant vectors. It is easy to check that $u_5$ and $u_6$ are the only invariant vectors of $A$ among the three levels. Also, $D$ has precisely $u_5$ and $u_6$ as invariant vectors among the three levels. However,
$A$ and $D$ are different in the sense that $A$ has order 4 and $D=A^2$ has order 2. For $P$ we have $Pv_2=v_2$, $Pv_3=v_3$ and $v_2,v_3$ are the only invariant vectors of $P$ among the three levels. Finally, $w_3,w_5$ are the only invariant vectors of
$L$ among the three levels. In summery, we have the following.

\begin{lem}       
\label{lem21}
Among the vectors in levels~1, 2 and 3, the elements of $\lscript$ have order 4 and precisely two invariant vectors in level~1, the elements of $\qscript$ have order 2 and precisely two invariant vectors in level~2, the elements of $\gscript$ have order 3 and precisely two invariant vectors in level~3 and the elements of $\bscript$ have order 2 and precisely two invariant vectors in level~1.
\end{lem}

It may be a coincidence that not counting the trivial identity element $I$, there are 23 elements in $\gscript _3$ which happens to be the number of basic elementary particles. However, it seems unlikely that such a coincidence would carry over to the fact that the elements of $\gscript _3$ band together into sets with common properties
\begin{equation*}
\gscript _3=\lscript\cup\qscript\cup\gscript\cup\bscript
\end{equation*}
and cardinalities $6,6,8,3$ the same size as the sets of leptons, quarks, gluons and bosons, respectively. In Section~3 we shall give other connections between these subsets of $\gscript _3$ and the elementary particles. The corresponding anti-particles can be treated by enlarging $\gscript _3$ to $\gscript '_3$. If $Z\in\gscript _3$ corresponds to a certain particle, then $-Z\in\gscript '_3$ corresponds to its anti-particle \cite{gudap2}.

The invariant vectors provide a method for visualizing the rotations in $\gscript _3$. We have seen that $u_5$ and $u_6$ are invariant vectors of $A\in\gscript _3$. We call the edge $u_5u_6$ an \textit{axis} and can visualize $A$ as a rotation of $\pi/2$ radians about the axis $u_5u_6$. Similarly, $D=A^2$ gives the rotation of $\pi$ radians about the axis $u_5u_6$. In this way the elements of $\lscript$ are rotations of $\pi/2$ about level~1 axes and the elements of $\bscript$ are rotations of $\pi$ about level~1 axes. In a similar way, the elements of $\qscript$ are rotations of $\pi$ about level~2 axes and elements of $\gscript$ are rotations $2\pi/3$ about level~3 axes \cite{gudap2}.

We call the set of invariant vectors together with the origin (or center) $c=(0,0,0)$ the \textit{elementary particle cube} $\cscript _3$. (Of course,
$Zc=c$ for all $Z\in\gscript _3$.) Thus,
\begin{equation}         
\label{eq21}
\cscript _3=\brac{c,u_1,\ldots ,u_6,v_1,\ldots ,v_{12},w_1,\ldots ,w_8}
\end{equation}
In this way, $\gscript  _3$ is the rotation group for the EPC $\cscript _3$. We think of $\cscript _3$ as the complete graph with vertices given in \eqref{eq21} and pairs of vertices such as $u_2v_7$ its edges. As with any graph, an edge $uv$ is considered to be the same as an edge $vu$ so edges are actually doubleton sets $\brac{u,v}$. We see that $\cscript _3$ has 27 vertices and $351=27\ctimes 26/2$ edges.

\section{Mass} 
We postulate that corresponding to each of the 23 elementary particles there is a unique complete subgraph of $\cscript _3$. That is, if $p$ is an elementary particle, then there is a corresponding set of vertices $\brac{x_1,x_2,\ldots ,x_n}\subseteq\cscript _3$ and their edges $x_1x_2,x_1x_3,\ldots ,x_{n-1}x_n$ that form a complete subgraph of the graph $\cscript _3$. We assume that the vertices represent tiny cells that are occupied by pre-elementary particles called preons and that the edges represent interactions between preons. The particular subgraph corresponding to an elementary particle will be determined by symmetry and the mass of the particle. In this approach, we view an elementary particle as having the shape of part of a cube which when its center of mass is translated to the origin $c$ it becomes a subgraph of $\cscript _3$.

We propose that the mass of a particle depends on the interactions between pairs of vertices $\brac{x,y}$ in its graph; that is, edges $xy$ in its graph. If $xy$ is an edge of $\cscript _3$, how do we find the interaction strength $s(x,y)$ of this edge? We have not found a method for deriving an equation for $s(x,y)$, but we do have some guidelines. First, $s(x,y)$ should be rotationally invariant; that is, $s(Zx,Zy)=s(x,y)$ for all $Z\in\gscript _3$. Second, it appears that the interactions could be similar to a  spring force that increases with distance. Now the main quantity that is rotation invariant is the norm
$\doubleab{x}$. Denoting the level by $\ell(x)=\doubleab{x}^2$, we postulate that $s(x,y)$ obeys the following ``fourth power law''
\begin{align}         
\label{eq31}
s(x,y)&=\sqbrac{2^{n(x,y)}\paren{\ell (x)+\ell (y)}}^2d(x,y)^4\notag\\
  &=\sqbrac{2^{n(x,y)}\paren{\doubleab{x}^2+\doubleab{y}^2}}^2\doubleab{x-y}^4
\end{align}
where
\begin{equation*}
n(x,y)=\begin{cases}1&\hbox{if }\ell (x)=\ell (y)=3\\0&\hbox{otherwise}\end{cases}
\end{equation*}
Equation~\eqref{eq31} states that $s(x,y)$ increases with the fourth power of the distance between $x$ and $y$, increases with the levels of $x$ and $y$ and kicks in a factor of 4 at the highest interaction levels $\ell (x)=\ell (y)=3$.

If $xy$ is an edge, we say that the \textit{interaction type} for $xy$ is $\ell (x)-\ell (y)$ where $\ell (x)\le\ell (y)$. For example, $cu_1$ has interaction type $0-1$, $v_1v_2$ has interaction type $2-2$ and $u_3w_5$ has interaction type $1-3$. We now compute some sample interaction strengths. Suppose we want $s(c,w_3)$. Since $n(c,w_3)=0$, $\ell (c)=0$, $\ell (w_3)=3$, $d(c,w_3)=\sqrt{3}$ we have
\begin{equation*}
s(c,w_3)=3^2(\sqrt{3})^4=81
\end{equation*}
To compute $s(v_1,w_2)$ we have $n(v_1,w_2)=0$, $\ell (v_1)=2$, $\ell (w_2)=3$ and $d(v_1,w_2)=\sqrt{5}$, Hence,
\begin{equation*}
s(v_1,w_2)=5^2(\sqrt{5})^4=625
\end{equation*}
Finally, to compute $s(w_1,w_2)$ we have $n(w_1,w_2)=1$, $\ell (w_1)=\ell (w_2)=3$ and $d(w_1,w_3)=2$ which gives
\begin{equation*}
s(w_1,w_3)=2^2\ctimes 6^2\ctimes 2^4=2304
\end{equation*}
Table~1 presents the interaction strengths for all the 351 edges of $\cscript_3$. The notation $c(u_1,u_2,\ldots ,u_6)$ means $cu_1,cu_2,\ldots cu_6$.


We postulate that the mass of an elementary particle is proportional to the sum of the interaction strengths for the edges of its corresponding graph plus a self-energy term. This self-energy term has the form $3n_p$ where $n_p=0$ if the graph for particle $p$ has fewer than 4 vertices and
$n_p$ is the number of vertices, otherwise. To set the mass units in MeV we take the proportionality constant to be the mass of the electron 0.511~MeV except for neutrinos where we use $(0.511)/3^4$~MeV. Thus, if $p$ is a neutrino, then its mass is given by 
\begin{align}         
\label{eq32}
m(p)&=\frac{(0.511)}{81}\,\sqbrac{\sum_{xy\in p}s(x,y)+3n_p}\\
\intertext{and otherwise,}
\label{eq33}        
m(p)&=(0.511)\sqbrac{\sum _{xy\in p}s(x,y)+3n_p}
\end{align}
where the notation specifies that $xy$ is an edge in the graph of $p$.
\eject

\thispagestyle{empty}
\vglue -4pc
{\parindent=-3pc
\begin{tabular}{c|c|c}
Type&Strength&Edges\hfill\hfill\\
\hline
$0-1$&1&$c(u_1,u_2,\ldots ,u_6)$\hfill\hfill\\\hline
$0-2$&16&$c(v_1,v_2,\ldots ,v_{12})$\hfill\hfill\\\hline
$0-3$&81&$c(w_1,w_2,\ldots ,w_8)$\hfill\hfill\\\hline
&&$u_1(u_3,u_4,u_5,u_6),u_2(u_3,u_4,u_5,u_6)$\hfill\hfill\\\
$1-1$&16&$u_3(u_5,u_6),u_4(u_5,u_6)$\hfill\hfill\\\cline{2-3}
&64&$u_1u_2,u_3u_4,u_5u_6$\hfill\hfill\\\hline
&&$u_1(v_1,v_3,v_5,v_7),u_2(v_2,v_4,v_6,v_8),u_3(v_1,v_2,v_9,v_{11})$\hfill\hfill\\
&9&$u_4(v_3,v_4,v_{10},v_{12}),u_5(v_5,v_6,v_9,v_{10}),u_6(v_1,v_8,v_{11},v_{12})$\hfill\hfill\\\cline{2-3}
$1-2$&&$u_1(v_9,v_{10},v_{11},v_{12}),u_2(v_9,v_{10},v_{11},v_{12}),u_3(v_5,v_6,v_7,v_8)$\hfill\hfill\\
&81&$u_4(v_5,v_6,v_7,v_8),u_5(v_1,v_2,v_3,v_4),u_6(v_1,v_2,v_3,v_4)$\hfill\hfill\\\cline{2-3}
&&$u_1(v_2,v_4,v_6,v_8),u_2(v_1,v_3,v_5,v_7),u_3(v_3,v_4,v_{10},v_{12})$\hfill\hfill\\
&225&$u_4(v_1,v_2,v_9,v_{11}),u_5(v_7,v_8,v_{11},v_{12}),u_6(v_5,v_6,v_9,v_{10})$\hfill\hfill\\\hline
&&$u_1(w_1,w_2,w_4,w_5),u_2(w_3,w_6,w_7,w_8),u_3(w_1,w_3,w_4,w_6)$\hfill\hfill\\
$1-3$&64&$u_4(w_2,w_5,w_6,w_7),u_5(w_1,w_2,w_3,w_7),u_6(w_4,w_5,w_6,w_8)$\hfill\hfill\\\cline{2-3}
&&$u_1(w_3,w_6,w_7,w_8), u_2(w_1,w_2,w_4,w_5),u_3(w_2,w_5,w_7,w_8)$\hfill\hfill\\
&576&$u_4(w_1,w_3,w_4,w_8),u_5(w_4,w_5,w_6,w_8),u_6(w_1,w_2,w_3,w_7)$\hfill\hfill\\\hline
&&$v_1(v_5,v_7,v_9,v_{11}),v_2(v_8,v_9,v_{11}),v_3(v_4,v_5,v_7,v_{10},v_{12})$\hfill\hfill\\
&64&$v_4(v_6,v_8,v_{10},v_{12}),v_5(v_9,v_{10}),v_6(v_9,v_{10}),v_7(v_{11},v_{12}),v_8(v_{11},v_{12})$\hfill\hfill\\\cline{2-3}
$2-2$&256&$v_1(v_2,v_3),v_2(v_4,v_6),v_5(v_6,v_7),v_8(v_6,v_7),v_9(v_{10},v_{11}),v_{12}(v_{10},v_{11})$\hfill\hfill\\\cline{2-3}
&&$v_1(v_6,v_8,v_{10},v_{12}),v_2(v_5,v_7,v_{10},v_{12}),v_3(v_6,v_8,v_9,v_{11})$\hfill\hfill\\
&576&$v_4(v_5,v_7,v_9,v_{11}),v_5(v_{11},v_{12}),v_6(v_{11},v_{12}),v_7(v_9,v_{10}),v_8(v_9,v_{10})$\hfill\hfill\\\cline{2-3}
&1024&$v_1v_4,v_2v_3,v_5v_8,v_6v_7,v_9v_{12},v_{10}v_{11}$\hfill\hfill\\\hline
&&$v_1(w_1,w_4),v_2(w_3,w_6),v_3(w_2,v_5),v_4(w_7,w_8),v_5(w_1,w_2),v_6(w_3,w_7)$\hfill\hfill\\
&25&$v_7(w_4,w_5),v_8(w_6,w_8),v_9(w_1,w_3),v_{10}(w_2,w_7),v_{11}(w_4,w_6),v_{12}(w_5,w_8)$\hfill\hfill\\\cline{2-3}
&&$v_1(w_2,w_3,w_5,w_6),v_2(w_1,w_4,w_7,w_8),v_3(w_1,w_4,w_7,w_8)$\hfill\hfill\\
$2-3$&&$v_4(w_2,w_3,w_5,w_6),v_5(w_3,w_4,w_5,w_7),v_6(w_1,w_2,w_6,w_8)$\hfill\hfill\\
&&$v_7(w_1,w_2,w_6,w_8),v_8(w_3,w_4,w_5,w_7),v_9(w_2,w_4,w_6,w_7)$\hfill\hfill\\
&625&$v_{10}(w_1,w_3,w_5,w_8),v_{11}(w_1,w_3,w_5,v_8),v_{12}(w_2,w_4,w_6,w_7)$\hfill\hfill\\\cline{2-3}
&&$v_1(w_7,w_8),v_2(w_2,w_5),v_3(w_3,v_6),v_4(w_1,w_4),v_5(w_6,w_8),v_6(w_4,w_5)$\hfill\hfill\\
&2025&$v_7(w_3,w_7),v_8(w_1,w_2),v_9(w_5,v_8),v_{10}(w_4,w_6),v_{11}(w_2,w_7),v_{12}(w_1,w_3)$\hfill\hfill\\\hline
&2304&$w_1(w_2,w_3,w_4),w_2(w_5,w_7),w_3(w_6,w_7),w_4(w_5,w_6),w_5w_8,w_6w_8,w_7w_8$\hfill\hfill\\\cline{2-3}
$3-3$&9216&$w_1(w_5,w_6,w_7),w_2(w_3,w_4,w_8),w_3(w_4,w_8),w_4w_8,w_5(w_6,w_7),w_6w_7$\hfill\hfill\\\cline{2-3}
&20736&$w_1w_8,w_2w_6,w_3w_5,w_4w_7$\hfill\hfill\\\hline
\hline\noalign{\medskip}
\multicolumn{3}{c}{\textbf{Table 1 (Interaction Strengths)}}
\end{tabular}
\parindent=18pt}

\noindent Equations \eqref{eq32}, \eqref{eq33} and a symmetry criterion essentially determine the graph of an elementary particle. Conversely, we shall show that the graph of an elementary particle determines its mass to within about $1\%$. The symmetry criterion, which is motivated by Lemma~\ref{lem21}, proceeds as follows. Lepton and boson graphs contain at least one level~1 vertex, quark graphs contain at least one level~2 vertex and gluon graphs contain at least one level~3 vertex.

\section{Graphs} 
To roughly describe the graph of an elementary particle, we  define the \textit{configuration} of a complete subgraph of $\cscript _3$ to be a 4-tuple
$(a_0,a_1,a_2,a_3)$ where $a_j$ is the number of vertices at level~$j$. By convention, $c=(0,0,0,0)$ is the only vertex at level~0. The configuration is invariant under the group $\gscript _3$ and indicates the geometry of the graph. However, it does not determine the graph uniquely. To accurately describe the graph, we must specify its vertices. Moreover, an isomorphic image under $\gscript _3$ will give the same mass. There are occasions when the mass formula does not uniquely determine the particle configuration. For example, the configurations $(1,0,1,0)$ and $(0,2,0,0)$ can both result in the mass 8.17~MeV. However, if this is to describe the down quark, then the second alternative is eliminated because a quark must have a level~2 vertex. Also, we shall see that $\nu _\tau$ and $\mu$ have the same configuration but different vertices and masses.

We now list our proposed configurations for the elementary particles. A gluon is massless so there are no interactions and it must therefore have just one vertex. Since this vertex must be at level~3, the gluon configuration is $g=(0,0,0,1)$. This is consistent with the fact that there are eight gluons and eight level~3 vertices.We propose that the electron neutrino has configuration $(0,1,0,0)$.This again gives zero mass which is consistent with its mass being less than $2.2\times 10^{-6}$~MeV, an amount that is orders of magnitude less than the other nonzero particle masses. Our list of configurations is the following:\medskip

{\obeylines
\noindent\textit{Neutrinos}:\enspace$\nu _e=(0,1,0,0)$, $\nu _\mu = (1,1,1,0)$, $\nu _\tau =(1,1,1,1)$
\noindent\textit{Leptons}:\enspace$e=(1,1,0,0)$, $\mu =(1,1,1,1)$, $\tau (1,1,2,1)$
\noindent\textit{Quarks}:\enspace$d=(1,0,1,0)$, $u=(0,1,1,0)$, $s=(1,1,2,0)$, $c=(0,1,2,1)$
\noindent\phantom{\textit{Quarks}:}\enspace$b=(1,2,2,2)$, $t=(1,4,12,8)$
\noindent\textit{Gluons}:\enspace$g=(0,0,0,1)$
\noindent\textit{Bosons}:\enspace$X^-=(1,4,12,5)$, $Z^0=(1,5,12,5)$, $H=(1,1,10,7)$
}\bigskip

Assuming these configurations, we now apply \eqref{eq32} and \eqref{eq33} for specific vertices to predict particle masses and compare them to experimental values which are given in parentheses.When considering experimental values, one must remember that neutrino masses have not been determined very precisely. This also applies to quark masses because quarks have not (cannot) be isolated so that their masses are essentially theoretical. One might argue that because of the flexibility of configurations and vertex choice, one can obtain practically any desired mass. However, with a closer examination, one can see that the mass values are fairly restricted. When applying \eqref{eq32}, \eqref{eq33}, we write the interaction strengths in the order of interaction type given in Table~1 followed by the $3n_p$ term.
\medskip

\noindent Muon neutrino: vertices $c,u_1,v_1$
\begin{equation*}
m(\nu _\mu )=\frac{(0.511)}{81}\,(1+16+9)=0.164\,(0.165)
\end{equation*}
Tau neutrino: vertices $c,u_1,v_2,w_2$
\begin{equation*}
m(\nu _\mu )=\frac{(0.511)}{81}\,(1+16+81+225+64+2025+12)=15.29\,(15.5)
\end{equation*}
Electron: vertices $c,u_1$
\begin{equation*}
m(e)=0.511\,(0.511)
\end{equation*}
Muon: vertices $c,u_1,v_1,w_1$
\begin{equation*}
m(\mu )=(0.511)\,(1+16+81+9+64+25+12)=106.29\,(105.6)
\end{equation*}
Tau: vertices $c,u_2,v_1,v_4,w_7$
\begin{align*}
m(\tau )&=(0.511)\,(1+32+81+234+64+1024+2050+15)\\
  &=1789.01\,(1776.8)
\end{align*}
Down Quark: vertices $c,v_1$
\begin{equation*}
m(d)=(0.511)\,(16)=8.17\,(\hbox{4-8})
\end{equation*}
Up Quark: vertices $u_1,v_1$
\begin{equation*}
m(u)=(0.511)\,(9)=4.599\,(\hbox{4-8})
\end{equation*}
Strange Quark: vertices $c,u_1,v_1,v_9$
\begin{equation*}
m(s)=(0.511)\,(1+32+90+64+12)=101.69\,(101)
\end{equation*}
Charm Quark: vertices $u_1,,v_{10},v_{11},w_5$
\begin{equation*}
m(c)=(0.511)\,(162+64+1024+1250+12)=1283.63\,(1270)
\end{equation*}
Bottom Quark: vertices $c,u_1,u_3,v_1,v_4,w_1,w_4$
\begin{align*}
m(b)&=(0.511)\,(2+32+162+16+468+256+1024+4100+2304+21)\\
  &=4284.74\,(4200)
\end{align*}
Top Quark: vertices $c,u_3,u_4,u_5,u_6,v_1,v_2,\ldots ,v_{12},w_1,w_2\ldots ,w_8$
\begin{align*}
m(t)&\!=\!(0.511)(4\!+\!192\!+\!648\!+\!320\!+\!5058\!+\!10240\!+\!24576\!+\!79200\!+\!221184\!+\!75)\\
  &\!=\!174505\,(173800)
\end{align*}
$X^-$ Boson: vertices $c,u_3,u_4,u_5,u_6,v_1,v_2,\ldots ,v_{12},w_4,w_5,w_6,w_7,w_8$
\begin{align*}
m(X^-)&\!=\!(0.511)(4\!+\!192\!+\!405\!+\!192\!+\!5040\!+\!6400\!+\!24576\!+\!49500\!+\!69120\!+\!66)\\
  &\!=\!79457.9\,(80,000)
\end{align*}
$Z^0$ Boson: vertices $c,u_2,u_3,\ldots ,u_6,v_1,v_2,\ldots ,v_{12},w_3,w_4,w_5,w_7,w_8$
\begin{align*}
m(Z^0)&\!=\!(0.511)(5\!+\!192\!+\!405\!+\!256\!+\!6300\!+\!8256\!+\!24576\!+\!49500\!+\!87552\!+\!69)\\
  &\!=\!90503.7\,(91,000)
\end{align*}
$H$ Boson: vertices $c,u_1,v_2,v_3,v_5,v_6,\ldots ,v_{12},w_2,w_3,\ldots ,w_8$
\begin{align*}
m(H)&\!=\!(0.511)\,(1+16+567+1026+2496+17600+58150+165888\!+\!57)\\
  &\!=\!125609\,(125,000)
\end{align*}

We point out that because of the three levels of vertices, this explains why leptons and quarks come in three generations.

\section{Composite Systems} 
We can apply our previous methods to determine masses of composite quark systems called hadrons, which consist of two types, the mesons and the baryons. As in the standard model, a meson is formed from a quark-antiquark pair
$q_1\qbar _2$ and a baryon is formed from a quark triple $q_1q_2q_3$. For the interactions between $q_1$ and 
$\qbar _2$, we do not consider the individual preon vertices of $q_1$ and $\qbar _2$ but view $q_1$ and $\qbar _2$ as single entities described by their configurations. Similarly, for a baryon $q_1q_2q_3$ we have interactions for the pairs
$(q_1,q_2)$, $(q_1,q_3)$ and $(q_2,q_3)$.

We propose that the mass formula for mesons is:
\begin{equation}         
\label{eq51}
m(q_1\qbar _2)=\tfrac{1}{2}\sqbrac{m(q_1)+m(q_2)}+(0.511)\sum s(x,y)
\end{equation}
where $x$ and $y$ are the noncentral (non-level 0) vertices of $q_1$ and $q_2$ and $s(x,y)$ are interaction strengths between $x$ and $y$ given in Table~1. We also propose that the mass formula for baryons is:
\begin{equation}         
\label{eq52}
m(q_1,q_2,q_3)=\tfrac{1}{3}\sqbrac{m(q_1)+m(q_2)+m(q_3)}+(0.511)\sum s(x,y)
\end{equation}
where $x,y$ are the noncentral vertices from pairs $(q_1,q_2)$, $(q_1,q_3)$, $(q_2,q_3)$ and $s(x,y)$ are again certain interaction strengths between $x$ and $y$ given in Table~1. The number of interactions of each type is determined by the configurations and the main problem is to decide which interaction strengths to choose from the various possibilities.

We postulate that in composite systems, the $u$ and $d$ quarks can interchange configurations. This is analogous to the standard model where, for example, a meson can be in a superposition of $u\ubar$ and $d\dbar$ mesons. We then define
\begin{equation*}
u'=(0,1,1,0)'=(1,0,1,0)
\end{equation*}
where $u'$ has a $d$ configuration, but retains mass $m(u)$ and the $u$ electric charge $(2/3)$. Similarly,
\begin{equation*}
d'=(1,0,1,0)'=(0,1,1,0)
\end{equation*}
where $d'$ has a $u$ configuration, but retains a mass $m(d)$ and the $d$ electric charge $(-1/3)$.

We now consider interaction strengths. In order to limit the number of choices of interaction strengths, we first reduce the possibilities in Table~1 to two or fewer representations. These are given in Table~2 in an order that we shall find convenient.

\begin{center}
\begin{tabular}{|c|c|c|c|c|c|c|}
\hline
Type&$1-1$&$1-2$&$2-2$&$1-3$&$2-3$&$3-3$\\
\hline
&16 (mesons)&&&&&\\
Strengths&64 (baryons)&\ 81,\ 225&\ 64,\ 256&\ 64,\ 576&\ 625,\ 2025&\ 2304,\ 9216\\
\hline\noalign{\medskip}
\multicolumn{7}{c}{\textbf{Table 2 (Representative Strengths)}}\\
\end{tabular}
\end{center}
\vskip 1pc

A hadron can be described by the configurations of its quark constituents. In turn, these configurations determine interactions of various types that we call the \textit{type sequences}
\begin{equation}         
\label{eq53}
\sqbrac{n_{11}(1-1),n_{12}(1-2),n_{22}(2-2),n_{13}(1-3),n_{23}(2-3),n_{33}(3-3)}
\end{equation}
where $n_{ij}$ is the number of type $(i-j)$ interactions for the quark constituents of the hadron. For example, the $K^+$ meson has the quark composition $K^+=u\sbar$. Since $u$ and $s$ have configurations $(0,1,1,0)$ and $(1,1,2,0)$ respectively, we write
\begin{equation*}
K^+=u\sbar =(0,1,1,0)(1,1,2,0)
\end{equation*}
Considering pairs of vertices in these two configurations, we see that there are one type $1-1$, three type $1-2$ and two type $2-2$ interactions. We then say that the type sequence for $K^+$ is
\begin{equation}         
\label{eq54}
\sqbrac{1(1-1),3(1-2),2(2-2)}
\end{equation}
For simplicity, we have omitted the zero terms from \eqref{eq53} in the expression \eqref{eq54}. For a more complicated example, the $\Lambda _c^+$ baryon has the composition
\begin{equation*}
\Lambda _c^+=udc=(0,1,1,0)(1,0,1,0)(0,1,2,1)
\end{equation*}
Considering the various pairs of vertices, we obtain the type sequence
\begin{equation*}
\sqbrac{1(1-1),5(1-2),5(2-2),1(1-3),2(2-3)}
\end{equation*}

We now define the generation of a hadron. The generation for mesons is similar to the usual generations of the quark constituents except now we have a quark-antiquark pair so there are more categories. The generation $g(q_1\qbar _2)$ is defined by
\begin{align*}
g(u\ubar )&=g(u\sbar )=2,\ g(s\sbar )=g(u\cbar )=4,\ g(s\cbar )=6,\ g(u\bbar )=7,\ g(c\cbar )=9\\
g(s\bbar )&=10,\ g(u\tbar )=14,\ g(c\bbar )=17,\ g(s\tbar )=18,\ g(c\tbar )=20,\ g(b\bbar )=26
\end{align*}
If a $d,u'$ or $d'$ appear instead of a $u$, the generation is the same as above. It will be useful later to introduce the \textit{generation number}
\begin{equation*}
\gamma (q_1\qbar _2)=6g(q_1\qbar _2)+\frac{\shat (1-\shat\,)}{2}+\bhat
\end{equation*}
where $\shat$ and $\bhat$ are the number of strange and bottom quark constituents, respectively for $q_1\qbar _2$.

The generation for baryons follows a similar pattern and we only list some of the lighter ones
\begin{equation*}
g(uud)=3,\ g(uus)=4,\ g(uss)=5,\ g(udc)=6,\ g(uuc)=g(sss)=7
\end{equation*}
Again, the generation does not change if $u$ or $d$ is replaced by $u'$ or $d'$. We also define the \textit{generation number}
\begin{equation*}
\gamma (q_1q_2q_3)=\begin{cases}7g(q_1q_2q_3)&\hbox{ if }n_{12}-n_{11}\le 4\\\noalign{\smallskip}
  7g(q_1q_2q_3)+\frac{6}{4-\shat}+\chat&\hbox{ otherwise}\end{cases}
\end{equation*}
\vskip 2pc
\noindent The next two tables summarize the lighter hadrons.
\bigskip

\hskip -2pc
\begin{tabular}{|c|c|c|c|c|c|c|}
\hline
&Quark&&&Type\\
Meson&Form&\;$\gamma$\;&Configuration&Sequence\\
\hline
$\pi ^0$&$u'\,\ubar '$&12&$(1,0,1,0)(1,0,1,0)$&$\sqbrac{1(2-2)}$\hfill\hfill\\
\hline
$K^+$&$u\,\sbar$&12&$(0,1,1,0)(1,1,2,0)$&$\sqbrac{1(1-1),3(1-2),2(2,2)}$\hfill\hfill\\
\hline
$\phi ^0$&$s\,\sbar$&23&$(1,1,2,0)(1,1,2,0)$&$\sqbrac{1(1-1),4(1-2),4(2-2)}$\hfill\hfill\\
\hline
$D^0$&$u\,\cbar$&24&$(0,1,1,0)(0,1,2,1)$&$\sqbrac{1(1-1),3(1-2),2(2-2),1(1-3),1(2-3)}$\\
\hline
$D_s^-$&$s\,\cbar$&36&$(1,1,2,0)(0,1,2,1)$&$\sqbrac{1(1-1),4(1-2),4(2-2),1(1-3),2(2-3)}$\\
\hline
$B^+$&$u\,\bbar$&43&$(0,1,1,0)(1,2,2,2)$&$\sqbrac{2(1-1),4(1-2),2(2-2),2(1-3),2(2-3)}$\\
\hline
$\eta _c^0$&$c\,\cbar$&54&$(0,1,2,1)(0,1,2,1)$&$\sqbrac{1(1-1),4(1-2),4(2-2),2(1-3),4(2-3)}$\\
\hline
$B_s^+$&$s\,\bbar $&61&$(1,1,2,0)(1,2,2,2)$&$\sqbrac{2(1-1),6(1-2),4(2-2),2(1-3),4(2-3)}$\\
\hline
\hline\noalign{\medskip}
\multicolumn{5}{c}{\textbf{Table 3 (Meson Structures)}}\\
\end{tabular}
\vskip 1pc

\hskip -4pc
\begin{tabular}{|c|c|c|c|c|c|c|}
\hline
&Quark&&&Type\\
Baryon&Form&\;$\gamma$\;&Configuration&Sequence\\
\hline
$P^-$&$uud'$&21&$(0,1,1,0)(0,1,1,0)(0,1,1,0)$&$\sqbrac{3(1-1),6(1-2),3(2-2)}$\hfill\hfill\\
\hline
$\Lambda ^0$&$uds$&28&$(0,1,1,0)(1,0,1,0)(1,1,2,0)$&$\sqbrac{1(1-1),5(1-2),5(2,2)}$\hfill\hfill\\
\hline
$\Sigma ^+$&$uus$&30&$(0,1,1,0)(0,1,1,0)(1,1,2,0)$&$\sqbrac{3(1-1),8(1-2),5(2-2)}$\hfill\hfill\\
\hline
$\Xi ^-$&$dss$&38&$(1,0,1,0)(1,1,2,0)(1,1,2,0)$&$\sqbrac{1(1-1),6(1-2),8(2-2)}$\hfill\hfill\\
\hline
$\Lambda _c^+$&$udc$&42&$(0,1,1,0)(1,0,1,0)(0,1,2,1)$&$\sqbrac{1(1\!-\!1),5(1\!-\!2),5(2\!-\!2),1(1\!-\!3),2(2\!-\!3)}$\\
\hline
$\Sigma _c^{++}$&$uuc$&50&$(0,1,1,0)(0,1,1,0)(0,1,2,1)$&$\sqbrac{3(1\!-\!1),8(1\!-\!2),5(2\!-\!2),2(1\!-\!3),2(2\!-\!3)}$\\
\hline
$\Omega ^-$&$sss$&55&$(1,1,2,0)(1,1,2,0)(1,1,2,0)$&$\sqbrac{3(1-1),12(1-2),12(2-2)}$\hfill\hfill\\
\hline
\hline\noalign{\medskip}
\multicolumn{5}{c}{\textbf{Table 4 (Baryon Structures)}}\\
\end{tabular}
\vskip 2pc

To apply \eqref{eq51} and \eqref{eq52} we choose interaction strengths from Table~2 according to the numbers given in the type sequence for the particular particle $p$. The resulting \textit{interaction path} for such a choice becomes
\begin{equation}         
\label{eq55}
s=\paren{n_0(p),n_1(p),\ldots n_{10}(p)}
\end{equation}
where $n_0$ is the number of 16s or 64s in type~$1-1$, $n_1$ is the number of 81s, $n_2$ is the number of 225s,\ldots ,
$n_{10}$ is the number of 9216s in the order of Table~2. There are many possible interaction paths for each hadron. For example,
$K^+$ has twelve possible interaction paths, some of which are
\begin{align}         
\label{eq56}
s_1&=(1,3,0,2,0,\ldots ,0)\notag\\
s_2&=(1,3,0,0,2,0,\ldots ,0)\\
s_3&=(1,3,0,1,1,0,\ldots ,0)\notag
\end{align}
Notice that the integers in the interaction path must conform with the integers in the type sequence. For example, in $s_3$ comparing with \eqref{eq54} we have $1+1=2$.

The \textit{interaction number} of an interaction path \eqref{eq55} for $p$ is
\begin{equation*}
n(s)=\sum _{j=0}^{10}jn_j(p)
\end{equation*}
An interaction path $s$ for a hadron $p$ is a \textit{mass path} for $p$ if $n(s)=\gamma (p)$. For example, the interaction numbers of the interaction paths in \eqref{eq56} are 9, 11, 10, respectively. Since $\gamma (K^+)=12$, none of these are mass paths for $K^+$.

If $s$ given in \eqref{eq55} is a mass path for a meson $p=q_1\qbar _2$, then applying \eqref{eq51} the corresponding
\textit{mass prediction} is
\begin{align}         
\label{eq57}
m(s)&=\tfrac{1}{2}\sqbrac{m(q_1)+m(q_2)}\notag\\
  &\qquad +(0.511)\sqbrac{16n_0(p)+81n_1(p)+\cdots +9216n_{10}(p)}
\end{align}
If $p=q_1q_2q_3$ is a baryon with mass path $s$ given in \eqref{eq55}, then applying \eqref{eq52} the corresponding mass prediction is
\begin{align}         
\label{eq58}
m(s)&=\tfrac{1}{3}\sqbrac{m(q_1)+m(q_2)+m(q_3)}\notag\\
  &\qquad +(0.511)\sqbrac{64n_0(p)+81n_1(p)+\cdots +9216n_{10}(p)}
\end{align}
For consistency, we employ the quark masses derived in Section~4. The \textit{mass} $m(p)$ of a hadron $p$ is defined to be the average of the mass predictions for the mass paths of $p$. That is, if $s_1,\ldots ,s_k$ are the mass paths for $p$ then
\begin{equation}         
\label{eq59}
m(p)=\frac{1}{k}\,\sum _{j=1}^km(s_j)
\end{equation}
In the exceptional case in which there does not exist a mass path for $p$ we take $m(p)=m(s)$ where $n(s)$ is closest to $\gamma (p)$.

We now consider two examples. The simplest case is the $\pi ^0$ meson which we propose has the quark composition
\begin{equation*}
\pi ^0=u'\ubar '=(1,0,1,0)(1,0,1,0)
\end{equation*}
The generation number is $\gamma (\pi ^0)=12$. Since there is only a single type $2-2$ interaction, the type sequence is
$\sqbrac{1(2-2)}$. We then have only the two interaction paths
\begin{align*}
s_1&=(0,0,0,1,0,\ldots ,0)\\
s_2&=(0,0,0,0,1,0,\ldots ,0)
\end{align*}
with $n(s_1)=3$, $n(s_2)=4$. As discussed earlier, since $n(s_2)=4$ is closest to $\gamma (\pi ^0)=12$, applying \eqref{eq57} we obtain
\begin{equation*}
m(\pi ^0)=m(s_2)=4.6+(0.511)256=135.42(135)
\end{equation*}
In a similar way, we have that
\begin{align*}
\pi ^+=u'\dbar&=(1,0,1,0)(1,0,1,0)\\
\intertext{and}
m(\pi ^+)&=6.38+(0.511)256=137.2(139.6)
\end{align*}
We treat $\pi ^-$ in a similar way. In the sequel, when we have charged versions of the same particle, we usually only treat one.

Our second example is the meson $K^+=u\sbar$ with type sequence \eqref{eq54} and generation number $\gamma (K^+)=12$. Now there are 12 interaction paths given by
\begin{align*}
s_1&=(1,3,0,2,0,\ldots ,0),\hskip 3pc s_2=(1,3,0,0,2,0,\ldots ,0)\\
s_3&=(1,3,0,1,1,0,\ldots ,0),\hskip 2pc s_4=(1,2,1,2,0,\ldots ,0)\\
s_5&=(1,1,2,2,0,\ldots ,0),\hskip 3pc s_6=(1,2,1,0,2,0,\ldots ,0)\\
s_7&=(1,1,2,0,2,0,\ldots ,0),\hskip 2pc s_8=(1,2,1,1,1,0,\ldots ,0)\\
s_9&=(1,1,2,1,1,0,\ldots ,0),\hskip 2pc s_{10}=(1,0,3,2,0,\ldots ,0)\\
s_{11}&=(1,0,3,0,2,\ldots ,0),\hskip 2.8pc s_{12}=(1,0,3,1,1,0,\ldots ,0)
\end{align*}
The interaction numbers for these paths are
\begin{align*}
n(s_1)&=9,\ n(s_2)=11,\ n(s_3)=10,\ n(s_4)=10,\ n(s_5)=11,\ n(s_6)=12,\\
n(s_7)&=13,\ n(s_8)=11,\ n(s_9)=12,\ n(s_{10})=12,\ n(s_{11})=14,\ n(s_{12})=13
\end{align*}
We conclude that $s_6$, $s_9$ and $s_{10}$ are mass paths. Applying \eqref{eq57} these give the following mass predictions:
\begin{align*}
m(s_6)&=\tfrac{1}{2}(4.6+101.69)+(0.511)(16+2\ctimes 81+225+2\ctimes 256)=520.71\\
m(s_9)&=\tfrac{1}{2}(4.6+101.69)+(0.511)(16+81+2\ctimes 225+64+256)=496.18\\
m(s_{10})&=\tfrac{1}{2}(4.6+101.69)+(0.511)(16+3\ctimes 225+2\ctimes 64)=471.65
\end{align*}
Applying \eqref{eq59} we obtain
\begin{equation*}
m(K^+)=\tfrac{1}{3}\sqbrac{m(s_6)+m(s_9)+m(s_{10})}=496.18(493.7)
\end{equation*}

\section{Predicted Masses} 
We now compute masses for the hadrons listed in Tables~3 and~4. We have already found masses for $\pi ^0$ and $K^+$ so we proceed for the others. All of these predicted masses agree with experiment to within $6\%$ and most of them agree much closer.

For the $\phi ^0$ meson we have $\gamma (\phi ^0)=23$ and mass paths
\begin{equation*}
s_1=(1,0,4,1,3,0,\ldots ,0),\quad s_2=(1,1,3,0,4,0,\ldots ,0)
\end{equation*}
This gives the mass predictions
\begin{align*}
m(s_1)&=101.69+(0.511)(16+4\ctimes 225+64+3\ctimes 256)=994.92\\
m(s_2)&=101.69+(0.511)(16+81+3\ctimes 225+4\ctimes 256)=1019.45
\end{align*}
The mass becomes
\begin{equation*}
m(\phi ^0)=\tfrac{1}{2}\sqbrac{m(s_1)+m(s_2)}=1007.18(1020)
\end{equation*}

For the $D^0$ meson, we have $\gamma (D^0)=24$ and the mass paths
\begin{align*}
s_1&=(1,0,3,2,0,1,0,1,0,\ldots ,0),\hskip 1.75pc s_2=(1,1,2,1,1,1,0,1,0,\ldots ,0)\\
s_3&=(1,1,2,2,0,0,1,1,0,\ldots ,0),\hskip 1.75pc s_4=(1,1,2,2,0,1,0,0,1,0,\ldots ,0)\\
s_5&=(1,2,1,0,2,1,0,1,0,\ldots ,0),\hskip 1.75pc s_6=(1,2,1,1,1,0,1,1,0,\ldots ,0)\\
s_7&=(1,2,1,1,1,1,0,0,1,0,\ldots ,0),\hskip 1pc s_8=(1,2,1,2,0,0,1,0,1,0,\ldots ,0)\\
s_9&=(1,3,0,0,2,0,1,1,0,\ldots ,0),\hskip 2pc s_{10}=(1,3,0,0,2,1,0,0,1,0,\ldots ,0)\\
s_{11}&=(1,3,0,1,1,0,1,0,1,0,\ldots ,0)
\end{align*}
This gives the mass predictions
\begin{align*}
m(s_1)&=1414.69,\hskip 1pc m(s_2)=1439.22,\hskip 1pc m(s_3)=1602.74\\
m(s_4)&=2056.5,\hskip 1.5pc m(s_5)=1463.74,\hskip 1pc m(s_6)=1627.26\\
m(s_7)&=2081.03,\hskip 1pc m(s_8)=2244.55,\hskip 1pc m(s_9)=1651.79\\
m(s_{10})&=2105.56,\hskip 1pc m(s_{11})=2269.08
\end{align*}
The mass becomes
\begin{equation*}
m(D^0)=\tfrac{1}{11}\sqbrac{m(s_1)+\cdots +m(s_{11})}=1814.2(1864.61)
\end{equation*}

For the $D_s^-$ meson, we have $\gamma (D_s^-)=36$ and the mass paths
\begin{align*}
s_1&=(1,3,1,4,0,1,0,2,0,\ldots ,0),\hskip 2pc s_2=(1,4,0,3,1,1,0,2,0,\ldots ,0)\\
s_3&=(1,4,0,4,0,0,1,2,0,\ldots ,0),\hskip 2pc s_4=(1,4,0,4,0,1,0,1,1,0,\ldots ,0)
\end{align*}
This gives the mass predictions
\begin{equation*}
m(s_1)=1742.25,\hskip 1pc m(s_2)=1766.78,\hskip 1pc m(s_3)=1930.3,\hskip 1pc m(s_4)=2499.05
\end{equation*}
The mass becomes
\begin{equation*}
m(D_s^-)=\tfrac{1}{4}\sqbrac{m(s_1)+m(s_{2})+m(s_3)+m(s_4)}=1984.6(1969)
\end{equation*}

For the $B^+$ meson, we have $\gamma (B^+)=43$ and the mass paths
\begin{align*}
s_1&=(2,0,4,0,2,0,2,1,1,0,\ldots ,0),\hskip 2pc s_2=(2,1,3,0,2,0,2,0,2,0,\ldots ,0)\\
s_3&=(2,0,4,1,1,0,2,0,2,0,\ldots ,0),\hskip 2pc s_4=(2,0,4,0,2,1,1,0,2,0,\ldots ,0)
\end{align*}
This gives the mass predictions
\begin{equation*}
m(s_1)=4825.38,\hskip 1pc m(s_2)=5467.19,\hskip 1pc m(s_3)=5442.65,\hskip 1pc m(s_4)=5279.13
\end{equation*}
The mass becomes
\begin{equation*}
m(B^+)=\tfrac{1}{4}\sqbrac{m(s_1)+m(s_{2})+m(s_3)+m(s_4)}=5253.59(5279)
\end{equation*}

For the $\eta _c^0$ meson, we have $\gamma (\eta _c^0)=54$ and there is only one mass path given by
\begin{equation*}
s=(1,4,0,4,0,2,0,4,0,\ldots ,0)
\end{equation*}
This gives the mass
\begin{align*}
m(\eta _c^0)&=m(s)=1283.63+(0.511)(16+4\ctimes 81+4\ctimes 64+2\ctimes 64+4\ctimes 625)\\
  &=2931.09(2983.6)
\end{align*}

For the $B_s^+$ meson, we have $\gamma (B_s^+)=61$. There are 35 mass paths which is too many to write down so we display a few:
\begin{align*}
s_1&=(2,1,5,4,0,2,0,4,0,\ldots ,0),\hskip 2pc s_2=(2,2,4,3,1,2,0,4,0,\ldots ,0)\\
s_{34}&=(2,6,0,3,1,2,0,0,4,0,\ldots ,0),\hskip 1pc s_{35}=(2,6,0,4,0,1,1,0,4,0,\ldots ,0)
\end{align*}
In summary, the mass becomes
\begin{equation*}
m(B_s^+)=\tfrac{1}{35}\sqbrac{m(s_1)+\cdots +m(s_{35})}=5366.53(5370)
\end{equation*}

We now consider baryons. The proton $P^+$ has $\gamma (P^+)=21$ and mass paths
\begin{align*}
s_1&=(3,0,6,3,0,\ldots ,0),\hskip 2pc s_2=(3,1,5,2,1,0,\ldots ,0)\\
s_3&=(3,2,4,1,2,0,\ldots ,0),\hskip 1pc s_4=(3,3,3,0,3,0,\ldots ,0)
\end{align*}
This gives the mass predictions
\begin{align*}
m(s_1)&=\tfrac{1}{3}(4.6+4.6+8.17)+(0.511)(3\ctimes 64+6\ctimes 225+3\ctimes 64)=891.86\\
m(s_2)&=\tfrac{1}{3}(4.6+4.6+8.17)+(0.511)(3\ctimes 64+81+5\ctimes 225+2\ctimes 64+256)\\
  &=916.39\\
m(s_3)&=\tfrac{1}{3}(4.6+4.6+8.17)+(0.511)(3\!\ctimes\!64+2\ctimes 81+4\ctimes 225+64+2\!\ctimes\!256)\\
  &=940.92\\
m(s_4)&=\tfrac{1}{3}(4.6+4.6+8.17)+(0.511)(3\ctimes 64+3\ctimes 81+3\ctimes 225+3\ctimes 256)\\
  &=965.45
\end{align*}
and mass
\begin{equation*}
m(P^+)=\tfrac{1}{4}\sqbrac{m(s_1)+\cdots +m(s_4)}=928.66(938.27)
\end{equation*}
The neutron $N^0=ud'd'$ is similar with mass
\begin{equation*}
m(N^0)=929.75(939.57)
\end{equation*}

The $\Lambda ^0$ baryon has $\gamma (\Lambda ^0)=28$ and mass paths
\begin{equation*}
s_1=(1,0,5,2,3,0,\ldots ,0),\ s_2=(1,1,4,1,4,0,\ldots ,0),\ s_3=(1,2,3,0,5,0,\ldots 0)
\end{equation*}
This gives the mass predictions
\begin{align*}
m(s_1)&=1103.59,\ m(s_2)=1128.12,\ m(s_3)=1152.64\\
\intertext{and mass}
m(\Lambda ^0)&=1128.12(1115.6)
\end{align*}

The $\Sigma ^+$ baryon has $\gamma (\Sigma ^+)=30$ and mass paths
\begin{align*}
s_1&=(3,1,7,5,0,\ldots ,0),\ s_2=(3,2,6,4,1,0,\ldots ,0),\ s_3=(3,3,5,3,2,0,\ldots ,0)\\
s_4&=(3,4,4,2,3,0,\ldots ,0),\ s_5=(3,5,3,1,4,0,\ldots ,0),\ s_6=(3,6,2,0,5,0,\ldots ,0)
\end{align*}
This gives the mass predictions
\begin{align*}
m(s_1)&=1144.81,\ m(s_2)=1169.34,\ m(s_3)=1193.87\\
m(s_4)&=1218.40,\ m(s_5)=1242.29,\ m(s_6)=1267.45
\intertext{and mass}
m(\Sigma ^+)&=1206.13(1189.4)
\end{align*}

The $\Xi ^-$ baryon has $\gamma (\Xi ^-)=38$ and mass paths
\begin{align*}
s_1&=(1,0,6,6,2,0,\ldots ,0),\ s_2=(1,1,5,5,3,0,\ldots ,0),\ s_3=(1,2,4,4,4,0,\ldots ,0)\\
s_4&=(1,3,3,3,5,0,\ldots ,0),\ s_5=(1,4,2,2,6,0,\ldots ,0),\ s_6=(1,5,1,1,7,0,\ldots ,0)\\
s_7&=(1,6,0,0,8,0,\ldots ,0)
\end{align*}
We then have the mass predictions
\begin{align*}
m(s_1)&=1250.93,\ m(s_2)=1274.50,\ m(s_3)=1299.98,\ m(s_4)=1324.51\\
m(s_5)&=1349.04,\ m(s_6)=1373.57,\ m(s_7)=1398.09
\intertext{and mass}
m(\Xi ^-)&=1324.51(1321.3)
\end{align*}

The $\Lambda _c^+$ baryon has $\gamma (\Lambda _c^+)=42$ and mass paths
\begin{align*}
s_1&=(1,2,3,5,0,1,0,2,0,\ldots ,0),\hskip 1.75pc s_2=(1,3,2,4,1,1,0,2,0,\ldots ,0)\\
s_3&=(1,3,2,5,0,0,1,2,0,\ldots ,0),\hskip 1.75pc s_4=(1,3,2,5,0,1,0,1,1,0,\ldots ,0)\\
s_5&=(1,4,1,3,2,1,0,2,0,\ldots ,0),\hskip 1.75pc s_6=(1,4,1,5,0,1,0,0,2,0,\ldots ,0)\\
s_7&=(1,4,1,4,1,0,1,2,0,\ldots ,0),\hskip  2pc s_8=(1,4,1,4,1,1,0,1,1,0,\ldots ,0)\\
s_9&=(1,4,1,5,0,0,1,1,1,0,\ldots ,0),\hskip 1.25pc s_{10}=(1,5,0,2,3,1,0,2,0,\ldots ,0)\\
s_{11}&=(1,5,0,5,0,0,1,0,2,0,\ldots ,0),\hskip 1.25pc s_{12}=(1,5,0,4,1,1,0,0,2,0,\ldots ,0)\\
s_{13}&=(1,5,0,4,1,0,1,1,1,0,\ldots ,0),\hskip 1.25pc s_{14}=(1,5,0,3,2,0,1,2,0,0,\ldots ,0)\\
s_{15}&=(1,5,0,3,2,1,0,1,1,0,\ldots ,0)
\end{align*}
This gives the mass predictions
\begin{align*}
m(s_1)&=1727.52,\ m(s_2)=1752.05,\ m(s_3)=1915.57,\ m(s_4)=2369.33\\
m(s_5)&=1776.57,\ m(s_6)=3011.15,\ m(s_7)=1940.09,\ m(s_8)=2393.86\\
m(s_7)&=2081.03,\ m(s_8)=2244.55,\ m(s_9)=1651.79,\ m(s_3)=1602.74\\
m(s_9)&=2557.38,\ m(s_{10})=1801.10,\ m(s_{11})=3199.20,\ m(s_{12})=3035.68\\
m(s_{13})&=2581.91,\ m(s_{14})=1964.62,\ m(s_{15})=2418.39
\end{align*}
Te mass becomes
\begin{equation*}
m(\Lambda _c^+)=\tfrac{1}{15}\sqbrac{m(s_1)+\cdots +m(s_{15})}=2296.29(2286)
\end{equation*}

The $\Omega ^-$ baryon has $\gamma (\Omega ^-)=55$ and mass paths
\begin{align*}
s_1&=(3,5,7,12,0,\ldots ,0),\ s_2=(3,6,6,11,1,0,\ldots ,0),\ s_3=(3,7,5,10,2,0,\ldots ,0)\\
s_4&=(3,8,4,9,3,0,\ldots ,0),\ s_5=(3,9,3,8,4,0,\ldots ,0),\ s_6=(3,10,2,7,5,0,\ldots ,0)\\
s_7&=(3,11,1,6,6,0,\ldots ,0),\ s_8=(3,12,0,5,7,0,\ldots ,0)
\end{align*}
We then have the mass predictions
\begin{align*}
m(s_1)&=1604.03,\ m(s_2)=1628.56,\ m(s_3)=1653.09,\ m(s_4)=1677.61,\\
m(s_5)&=1702.14,\ m(s_6)=1726.67,\ m(s_7)=1751.20,\ m(s_8)=1775.73
\end{align*}
The mass becomes
\begin{equation*}
m(\Omega ^-)=\tfrac{1}{8}\sqbrac{m(s_1)+\cdots +m(s_8)}=1689.88(1672.5)
\end{equation*}

Our last example is the $\Sigma _c^{++}$ baryon with $\gamma (\Sigma _c^{++})=50$ and mass paths
\begin{align*}
s_1&=(3,5,3,5,0,2,0,2,0,\ldots ,0),\hskip 1.75pc s_2=(3,6,2,4,1,2,0,2,0,\ldots ,0)\\
s_3&=(3,6,2,5,0,1,1,2,0,\ldots ,0),\hskip 1.75pc s_4=(3,6,2,5,0,2,0,1,1,0,\ldots ,0)\\
s_5&=(3,7,1,3,2,2,0,2,0,\ldots ,0),\hskip 1.75pc s_6=(3,7,1,5,0,0,2,2,0,\ldots ,0)\\
s_7&=(3,7,1,5,0,2,0,0,2,\ldots ,0),\hskip  1.75pc s_8=(3,7,1,4,1,1,1,2,0,\ldots ,0)\\
s_9&=(3,7,1,4,1,2,0,1,1,0,\ldots ,0),\hskip .7pc s_{10}=(3,7,1,5,0,1,1,1,1,0,\ldots ,0)\\
s_{11}&=(3,8,0,2,3,2,0,2,0,\ldots ,0),\hskip 1.75pc s_{12}=(3,8,0,3,2,1,1,2,0,\ldots ,0)\\
s_{13}&=(3,8,0,4,1,0,2,2,0,\ldots ,0),\hskip 1.75pc s_{14}=(3,8,0,3,2,2,0,1,1,0,\ldots ,0)\\
s_{15}&=(3,8,0,4,1,2,0,0,2,0,\ldots ,0),\hskip 1pc s_{16}=(3,8,0,5,0,0,2,1,1,0,\ldots ,0)\\
s_{17}&=(3,8,0,5,0,1,1,0,2,0,\ldots ,0),\hskip 1.25pc s_{18}=(3,8,0,4,1,1,1,1,1,0,\ldots ,0)\\
\end{align*}
This gives the mass predictions
\begin{align*}
m(s_1)&=1948.61,\ m(s_2)=1973.41,\ m(s_3)=2136.66,\ m(s_4)=2590.43\\
m(s_5)&=1997.67,\ m(s_6)=2324.71,\ m(s_7)=3232.25,\ m(s_8)=2161.19\\
m(s_9)&=2614.96,\ m(s_{10})=2778.48,\ m(s_{11})=2022.20,\ m(s_{12})=2185.72\\
m(s_{13})&=2349.24,\ m(s_{14})=2639.49,\ m(s_{15})=3256.77,\ m(s_{16})=2966.53\\
m(s_{17})&=3420.29,\ m(s_{18})=2803.01
\end{align*}
The mass becomes
\begin{equation*}
m(\Sigma _c^{++})=\tfrac{1}{18}\sqbrac{m(s_1)+\cdots +m(s_{18})}=2522.31(2452)
\end{equation*}

We close with two remarks. First, we have not considered electric forces in this work. The strengths that we introduced evidently correspond to the weak and strong nuclear forces. Presumably, the neglect of electric forces is responsible for the small discrepancies in our mass computations. Second, we have only presented basic hadrons and have not considered excited states or resonances. For example
\begin{equation*}
\omega ^0=u\ubar =(0,1,1,0(0,1,1,0)
\end{equation*}
is a resonances with interaction sequence
\begin{equation*}
\sqbrac{1(1-1),2(1-2),1(2-2)}
\end{equation*}
we have that $\gamma (\omega ^0)=12$ and the interaction path
\begin{equation*}
s=(1,0,2,0,1,0,\ldots ,0)
\end{equation*}
has interaction number $n(s)=8$ which is the closest to $\gamma (\omega ^0)$. The usual mass prediction $m(\omega ^0)=m(s)$ is much smaller than the experimental value. However, if we employ other interaction strengths from Table~1 we can obtain
\begin{equation*}
m(\omega ^0)=4.6+(0.511)(64+2\ctimes 225+1024)=790.5(782)
\end{equation*}
This indicates that more flexibility is required if resonances are to be included.



\begin{thebibliography}{99}
\bibitem{bdp16}A.\,Bisco, G.\,D'Ariano and P.\,Perinotti, Special relativity in a discrete quantum universe, arXiv: quant-ph 1503.01017v3 (2016).
\bibitem{cro16}D.\,Crouse, On the nature of discrete space-time, arXiv: quant-ph 1608.08506v1 (2016).
\bibitem{gud68}S.\,Gudder, Elementary length topologies in physics, \textit{SIAM J.\ Appl.\ Math.}\textbf{16} 1011--1019 (1968).
\bibitem{gud161}--------, Discrete quantum gravity and quantum field theory, arXiv: gr-qc 1603.03471v1 (2016).
\bibitem{gud162}--------, Discrete scalar quantum field theory, arXiv: physics.gen-ph 1610.07877v1 (2016).
\bibitem{gudap1}--------, Reconditioning in discrete quantum field theory, \textit{Int.\ J.\ Theor.\ Phys.} (to appear).
\bibitem{gudap2}--------, Discrete spacetime and quantum field theory, arXiv: physics.gen-ph 1704.01639 (2017).
\bibitem{hei30}W.~Heisenberg, \textit{The Physical Principles of Quantum Mechanics}, University of Chicago Press, Chicago (1930).
\bibitem{rus54}B.~Russell, \textit{The Analysis of Matter}, Dover, New York (1954).

\end{thebibliography}
\end{document}